\RequirePackage{amsmath}
\documentclass{svproc}
\usepackage{amsmath}
\usepackage{multirow}
\usepackage{makecell}
\usepackage{array}
\usepackage{enumitem}
\usepackage{graphicx}

\usepackage{url}

\usepackage[T1]{fontenc}
\usepackage{lmodern}
\usepackage{microtype}
\DisableLigatures[f]{encoding = *, family = tt* }

\usepackage{color}
\PassOptionsToPackage{hyphens}{url}\usepackage{hyperref}
\hypersetup{hidelinks}

\begin{document}
\mainmatter              
\title{Safety-centric and Smart Outdoor Workplace: A New Research Direction and Its Technical Challenges\thanks{Supported in part by Chilean National Research and Development Agency (ANID, Chile) under Grant
FONDECYT Iniciaci{\'o}n 11180905.}}
\titlerunning{Safety-centric and Smart Outdoor Workplace}  
%
\author{Zheng Li\inst{1}$^{\texttt{[ORCID: 0000-0002-9704-7651]}}$ \and
Mauricio Pradena Miquel\inst{2} \and
Pedro Pinacho-Davidson\inst{1}}
\authorrunning{Zheng Li et al.} 
%
%
\institute{Department of Computer Science, University of Concepci{\'o}n, Concepci{\'o}n, Chile\\
\email{\{zli,ppinacho\}@udec.cl} \and
Department of Civil Engineering, University of Concepci{\'o}n, Concepci{\'o}n, Chile\\
\email{mpradena@udec.cl}}

\maketitle              

\begin{abstract}
Despite the fact that outside is becoming the frontier of indoor workplaces, a large amount of real-world work like road construction has to be done by outdoor human activities in open areas. Given the promise of the smart workplace in various aspects including productivity and safety, we decided to employ smart workplace technologies for a collaborative outdoor project both to improve the work efficiency and to reduce the worker injuries. Nevertheless, our trials on smart workplace implementation have encountered a few problems ranging from the theoretical confusion among different stakeholders, to the technical difficulties in extending underground devices' lifespan. This triggers our rethinking of and discussions about ``smart workplace''. Eventually, considering the unique characteristics of outdoor work (e.g., more sophisticated  workflows and more safety-related situations than office work), we argue that ``safety-centric and smart outdoor workplace'' deserves dedicated research attentions and efforts under the umbrella discipline of smart environment. In addition, the identified technical challenges can in turn drive different research dimensions of such a distinguishing topic.
\end{abstract}

\section{Introduction}
\label{sec:introduction}

Outside place is recently claimed to be the new workplace frontier \cite{Schneider_2016}, whereas this opinion essentially treats outside as an extension of the indoor workplace. In fact, numerous types of real-world work have been extensively relying on outdoor human activities (e.g., road construction). Compared with working in the indoor places, outdoor workers inevitably encounter more environmental impacts and more safety risks, ranging from sunburns to accidents. Therefore, it would be more meaningful and crucial to make outdoor workplaces smart to reduce injuries and health threats, in addition to pursuing the other benefits like increasing productivity and seamless communication. 

Following the naming convention, ``smart outdoor workplace'' belongs to the area of smart workplace, while smart workplace is a specific type of smart environment that can be traced back to the late 1980s when ubiquitous computing emerged \cite{Weiser_Gold_1999}. In general, the ultimate goal of any smart environment is to improve its inhabitants' experience in that environment \cite{Cook_Das_2005,Maestre_Lopez_2006}. Along the evolution of ubiquitous computing, smart environment keeps booming as a hot topic in both academia and industry. For example, when it comes to Internet of Things (IoT) as the modern paradigm of ubiquitous computing, one of its major application areas is still Smart Environment IoT \cite{Curry_Sheth_2018,Gubbi_Buyya_2013}. As a matter of fact, the relevant enabling technologies are turning the influential vision of the physical world \cite{Weiser_1991} into reality, i.e.~the omnipresent computational devices, sensors, actuators and other elements are being substantially and intelligently interwoven together to promote quality of life and to assist human in everyday tasks at various kinds of environments \cite{Kbar_Aly_2014}. 

Among different smart environments, smart workplace has attracted considerable attentions aiming to facilitate decision making, enhance job efficiency, automate tedious routine, and also improve working comfort and happiness \cite{Augusto_2010,Mikulecky_2012}. According to a recent survey, 72 percent of enterprise organizations have introduced IoT devices and sensors into the workplace, ranging from air conditioning and lighting systems (56 percent) to personal mobile devices (51 percent) \cite{Gibson_2017}. By improving effectiveness of employees and enabling them to contextually interact with connected ``things'', the IoT-supported workplace and environments have been considered smart to fuel better ways of doing business. According to Gartner's report on digital government technology, smart workplace is approaching the peak of inflated expectations \cite{Memoori_2018,Goasduff_2019}. In practice, many enterprises have improved the effectiveness of their IT teams and increased profitability through smart workplace implementation. When prospecting the next generation of technologies, smart workplace even represents a significant technology trend towards transparently immersive experiences \cite{Gartner_2018}.

When we tried to implement smart workplace for a collaborative project with Chile's national highway agency, unfortunately, we met various issues and challenges in both theory and practice due to the unique outdoor characteristics. Therefore, we decided to systematically summarize those issues and challenges together with potential research prospects, to both facilitate guiding our own project and help boom the community of smart outdoor workplace. This paper reports our most recent effort. To the best of our knowledge, this is the first study distinguishing smart outdoor workplace from the umbrella discipline of smart environment. 

The remainder of this paper is organized as follows. Section \ref{sec:need} justifies the need of dedicated focus on safety-centric and smart outdoor workplace. Section \ref{sec:issues} highlights the main technical challenges in studying and implementing smart outdoor workplace. Section \ref{sec:directions} correspondingly suggests a set of research prospects driven by the identified research challenges. Conclusions and some future work are discussed in Section \ref{sec:conclusion}.

\section{The Need of Dedicated Focus on Smart Outdoor Workplace}
\label{sec:need}
After the term ``smart environment'' was coined in the late 1980s, there came increasing interests in and efforts on various smart environment forks. For example, the early definition of smart home appeared with the influence of home automation terminology in 1992 \cite{Lutolf_1992,Alam_Reaz_2012}; while the smart city term emerged in literature from urban simulations and knowledge bases in 1998 \cite{van_Bastelaer_1998,Anthopoulos_2015}. Unfortunately, due to the overlap of supporting technologies and application scenarios, the meaning and context of different smart environments could still be confusing \cite{Anthopoulos_Fitsilis_2013}. To our best knowledge, there is particularly a lack of scientific landscaping work to facilitate understanding the specific domain of smart workplace, which in turn drives the need of clear and dedicated focus on safety-centric and smart outdoor workplace.

\subsection{Smart Workplace}
Given the layered architecture of various smart environments, smart workplace is generally considered as a comparable counterpart to other human-centric surroundings like smart home under the umbrella concept of smart city \cite{Maestre_Lopez_2006,Georgantas_Issarny_2010}; however, smart workplace can also become a subdomain of smart home, e.g., in the case of working from home \cite{Cho_Kim_2013}. When it comes to its own definition, smart workplace has come with different aliases and sometimes inconsistent substances. For example, smart workplace widely refers to smart office and smart/intelligent room \cite{Mikulecky_2012}, whereas it has also been extended to include smart building \cite{iSCOOP_2018}. In addition, this term may suffer from a frequent semantic confusion, i.e.~the alternately-used ``smart workplace'' with ``smart workspace'' even in the same articles (e.g., \cite{Cho_Kim_2013,Memoori_2018}). Rigorously speaking, workplace and workspace can comprise distinct essences for representing where and how work happens respectively, and therefore workspace has been treated as a component of workplace \cite{Bicknell_2017}.

\subsection{Smart Outdoor Workplace}

\begin{figure}[!t]
	\centering
	\includegraphics[width=8.5cm,trim=325 312 320 165,clip]{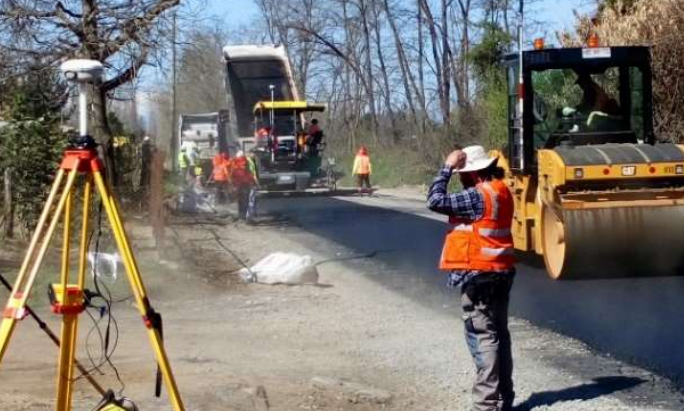}    
	\caption{Road construction workplace is a typical example of outdoor workplaces.}
	\label{figPave}
\end{figure}

In most cases, the same type of specific environments (e.g., home or restaurant \cite{Maestre_Lopez_2006}) is composed of common elements to support common routines. However, workplace environments could be vague and difficult to unify, because different jobs at different places can involve completely different human activities, which might be the reason that much less efforts have been invested into smart workplace compared to the other smart environments like smart home \cite{Rocker_2010}. Although it is impossible to outline a generic working environment, we can roughly distinguish between indoor and outdoor workplaces, and they do not always share the same purposes. For example, to improve efficiency and productivity, smart indoor workplace could focus more on individual employees' separation and privacy \cite{Cho_Kim_2013}, while smart outdoor workplace (e.g., the road construction workplace as illustrated in Fig.~\ref{figPave}) would emphasize real-time cooperation among various roles \cite{Pradena_Miller_2019}. Furthermore, since outdoor workplaces are often linked with death and injuries to workers and members of the public, safety in smart outdoor workplace must be a crucial concern as stressed by Safe Work Australia \cite{SWA_2014}.

\subsection{Divergent Technology Streams} 
Smart workplace is associated with at least two main technology streams, in the contexts of the traditional ubiquitous computing and the modern edge computing respectively. On one hand, it has been widely accepted that smart environment evolves through exploiting the key enabling technologies from ubiquitous computing together with Ambient Intelligence (AmI) that aims to make an environment behave intelligently \cite{Augusto_2010,Kbar_Aly_2014}. For example, smart home and smart workplace are highlighted as two types of applications of ubiquitous computing \cite{Alam_Reaz_2012} and AmI \cite{Mikulecky_2012}. In particular, the implementation of AmI is because it provides a sensible assistance considering the user preferences, mood, and the current overall situation for increase situational awareness \cite{Augusto_2007}, thus reducing distraction and the consequent injury risks. On the other hand, based on the ambitious vision of a Network of pervasively interconnected objects \cite{Greengard_2015}, smart environment has been claimed to be the major application area of IoT, although it still lacks well-established IoT-driven practices due to the novelty and complexity \cite{Zanella_Bui_2014}. Benefiting from the most advanced communication technologies and especially the release of 5G, ultra-low-latency edge computing receives tremendous interests and development \cite{Satyanarayanan_2017}, which in turn fosters the wide deployment and application of IoT \cite{Froehlich_2018,Premsankar_Francesco_2018,Salman_Elhajj_2015}. As a result, implementing IoT-enabled smart workplace has become a clear trend and become more promising than ever in practice \cite{Esparza_2018,Joshi_2018,OSSAZ_2018}.

~\\

Overall, it is necessary and beneficial to clarify and characterize different environmental scopes by using standard terminology, integrate the relevant research efforts by fusing different technology streams, and eventually solid a research foundation by portraying out the landscape of the smart outdoor workplace domain.

\section{Technical Challenges}
\label{sec:issues}
According to our experience in and lessons from the ongoing project, two typical challenges seem to be addressing the diversity in outdoor devices (over the off-the-shelf choices) and extending their power life (especially in the case of underground installation). Note that the challenges in the social and environmental contexts are not discussed in this paper, as they are out of the scope of technical concerns at this current stage. 

\subsection{Diverse and Incompatible Sensors and Devices}
Computing devices have experienced a revolutionary evolution from large mainframes to small chips that can be embedded in a variety of equipment and places. This leads to an exploding diversity in physical components involved in any smart environment, not to mention different workplace scenarios.  When implementing a smart workplace in any real-world scenario, it is undoubtedly vital to effectively employ various physical components \cite{Cook_Das_2007} such as sensors, controllers, and wearable devices that could have been silently distributed both in public spaces and in our more private surroundings \cite{Mikulecky_2012}.

Unfortunately, the sensor industry has not employed any globally standard protocols and universal interfaces yet, which is similar to the early computer industry \cite{Gorman_Resseguie_2009}. On the contrary, to secure the niche market shares against competitors, different sensor vendors even tend to keep developing their own proprietary standards and continue the existing incompatible protocols. Consequently, the main challenges in smart workplace implementations are related to the non-scalable integration of heterogeneous technologies produced by different sensor/device manufacturers \cite{Augusto_2010,Viani_Robol_2013}. To address these challenges, following the software-driven trend in making computing environment programmable and software defined \cite{Li_Brech_2014}, smart workplace (and any smart environment) should also pay more attentions to software to deal with the possibly tremendous heterogeneity and diversity in the sensor technologies.

More importantly, despite of the widely-recognized and possibly-overemphasized hardware aspects  \cite{Cook_Das_2007}, it is software that eventually makes a whole workplace smart. Inspired by the well known ``mind/brain'' metaphor \cite{Sachs_2016}, users essentially rely on software applications to communicate and work with smart devices and objects around them. In particular, it has been revealed that ``the success in the industrial sector where data and communication equate to lost lives and billions of dollars largely depends on software's ability to create valuable functionality'' \cite{Chasty_2013}.

Overall, given the software characteristics of all smart environments, we consider smart workplace to be part of the software-driven world \cite{Bosch_2016}. Thus, promising solutions to this hardware-side research problem would still rest with software aspects and data integration, i.e.~software-driven smart (outdoor) workplace. 

\subsection{Growing Gap between Battery Technologies and Power Requirements}

\begin{figure}[!t]
	\centering
	\includegraphics[width=8.5cm,trim=0 0 0 0,clip]{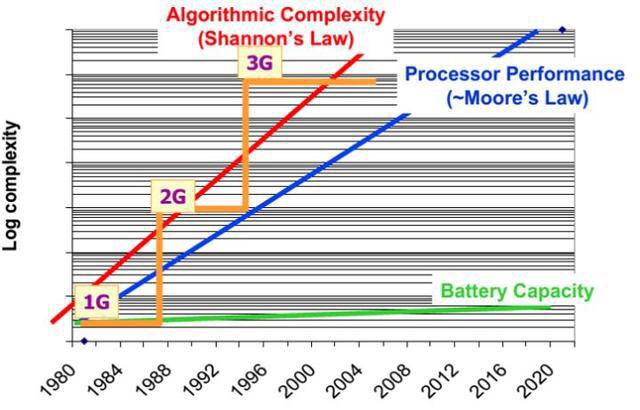}    
	\caption{Eveready's Law: Battery power grows much more slowly than the requirement of consuming it \cite{Leibson_2006}.}
	\label{figBattery}
\end{figure}

Due to the inherent difficulties and complexity in the relevant interdisciplinary topics (e.g., thermodynamics and fluid mechanics), there is a tremendously growing gap between battery technology and power requirements (cf.~Eveready's Law in Fig.~\ref{figBattery}). As a result, limited energy supply could be a critical constraint for the pervasive battery-powered devices/sensors in a generic sense. More crucially, in special environments such as underwater, underground, or within buildings, energy supply has been identified as a major challenge in wireless sensor networks, because the involved sensors or devices are unable to be recharged after installation \cite{Aziz_2013}. Thus, research and practice in any smart environment need inevitably to take into account reducing power/energy consumption as well as maintaining the functionality.

Achieving the goal of ultra-low power/energy consumption requires multi-level efforts ranging from the system architecture to the circuit technology \cite{Rabaey_Ammer_2002}. Since the current progress in both physics and nano-electronics has not yet broken through the power bottleneck of conventional MOS-based integrated circuits \cite{Kang_Zhang_2015}, it can be seen that more power-dissipation efforts are turning towards software-driven solutions, such as topology control techniques that algorithmically enable sensor nodes to control certain parameters to extend the lifetime of a wireless sensor network \cite{Aziz_2013}.  
Meanwhile, the public has also started being aware of software energy consumption especially in the context of personal devices. For example, due to some problematic features of the operating software iOS 11, many iPhone users have suffered from shorter and shorter full charge cycles since their last OS upgrade, and a suggested workaround is to get into the software and uninstall some apps who ``kill'' battery life the most \cite{Smith_2017}.

Similar to the limited commentary regarding software's vital role in smart workplace \cite{Chasty_2013}, unfortunately, the existing concerns and efforts mainly stress the importance of improving the energy efficiency of infrastructural IT equipment and smart devices (e.g., \cite{Bento_2016,Piyare_Murphy_2017,Terry_Palmer_2016}). Practitioners generally lack consistent knowledge regarding the energy consumption relationship between software and hardware \cite{Pang_Hindle_2016}, not to mention the lack of models, descriptions, or realizations of software energy consumption in smart environments. Note that the aforementioned topology control techniques could still lead to local energy optimization without the holistic context of a software application. In other words, sustainability has not played a major role in the development of smart software systems, even though reducing energy consumption has been argued to be the first-class target over performance goals when software is being designed since two decades ago \cite{Ellis_1999}.

Recall that there are hardly any systematic methods available that try to integrate sustainability aspects into software systems, as it's common today for material products like cars, light bulbs or computer hardware \cite{Kern_2013}. In addition to the study on software-driven smart (outdoor) workplace, it will also be significantly crucial and beneficial to investigate the energy consumption foundations of, and improve the energy efficiency of, software systems within smart (outdoor) workplace. 

\section{Research Prospects Driven by the Identified Challenges}
\label{sec:directions}
Given the previous discussions, we suggest developing a knowledge architecture, a microservice-oriented platform, and a plug-and-play modeling framework to address the aforementioned research gaps and challenges respectively. In particular, we have initially investigated the microservice-oriented software solution to deal with the device heterogeneity in a proof-of-concept study \cite{Li_Seco_2019}. To avoid duplication, here we only highlight the other two research prospects in this paper.

\subsection{Sandwich Approach to a Knowledge Architecture for Smart Outdoor Workplace}
Considering that a generic feature of \textit{Knowledge Architecture} is being able to reduce confusion in human knowledge, we can used a domain-specific knowledge architecture not only to clarify the landscape of, but also to facilitate reusing the relevant knowledge of, smart outdoor workplace. Here we adopt NASA's definition that treat knowledge architecture as a combination of information architecture, knowledge management, and data science \cite{Meza_2016}, as specified in Equation (\ref{eq_KA}). Note that the detailed explanation about \textit{Knowledge Architecture} is out of the scope of this paper. 
\begin{equation}
\label{eq_KA}
\begin{split}
\textit{Knowledge Architecture} ~=~&\textit{Information Architecture}~+\\
                  &\textit{Knowledge Management}~+\\
                  &\textit{Data Science}
\end{split}
\end{equation}%

Following Equation (\ref{eq_KA}), we suggest employing a sandwich approach to develop a domain-specific knowledge architecture, by collecting and investigating both theoretical data and practical data. In particular, inspired by the generic factor-based system model \cite{Montgomery_2012},  the relevant knowledge can be organized by distinguishing between input factors, environmental factors, and response (output) factors within outdoor workflows. 
\begin{itemize}
\setlength\itemsep{1em}
\item \textbf{Theoretical Investigation} should focus on the state-of-the-art of the smart workplace domain. We suggest employing an evidence-based research method to study the existing theoretical discussions. In fact, evidence-based research has been advocated as a prerequisite to new studies across a wide range of disciplines \cite{Kitchenham_2015_book,Lund_2016}. By identifying relevant studies, summarizing previous findings, and comparing the related work, we can obtain a rough structure of the knowledge architecture for smart workplace in general (i.e.~not limited to outdoor workplace). 

\item \textbf{Empirical Investigation} should be based on real-world outdoor work projects. We are currently employing the scenario of road construction to exemplify outdoor workplace (cf.~the rough workflow breakdown in Table \ref{table_breakdown}). By attracting more attentions and by involving more real-world cases, we expect to identify the characteristics, dimensions and elements of outdoor workplace (as the counterpart of and the supplement to the theoretical investigation), and then gradually make the proposed knowledge architecture concrete and dedicated to smart outdoor workplace.
\end{itemize}

\begin{table}[!t]\footnotesize
\renewcommand{\arraystretch}{1.3}
\caption{A Rough Breakdown of the Workflow and Measurable Data in Road Construction Projects \cite{PAIKY_2012}}
\label{table_breakdown}
\centering
\begin{tabular}{>{\raggedright}p{1.3cm} >{\raggedright}p{2.4cm} >{\raggedright}p{2.5cm} >{\raggedright}p{2.6cm} >{\raggedright\arraybackslash}p{2.6cm}}
\hline

\hline
& \textbf{Transporting hot mix asphalt} & \textbf{Discharging asphalt into paver}& \textbf{Creating mat and initial compaction}& \textbf{Compaction and quality control}\\
\hline
\textbf{Past} &
    \begin{itemize}[nolistsep,leftmargin=*,before*={\mbox{}\vspace{-\baselineskip}},after*={\mbox{}\vspace{-0.7\baselineskip}}]
        \item 	N/A.
    \end{itemize}
 & 
    \begin{itemize}[nolistsep,leftmargin=*,before*={\mbox{}\vspace{-\baselineskip}},after*={\mbox{}\vspace{-0.7\baselineskip}}]
        \item 	N/A.
    \end{itemize} 
 & 
    \begin{itemize}[nolistsep,leftmargin=*,before*={\mbox{}\vspace{-\baselineskip}},after*={\mbox{}\vspace{-0.7\baselineskip}}]
        \item 	Previous paths.
    \end{itemize} 
 & 
    \begin{itemize}[nolistsep,leftmargin=*,before*={\mbox{}\vspace{-\baselineskip}},after*={\mbox{}\vspace{-0.7\baselineskip}}]
        \item 	Previous paths.
    \end{itemize} \\

\textbf{Present} &
    \begin{itemize}[nolistsep,leftmargin=*,before*={\mbox{}\vspace{-\baselineskip}},after*={\mbox{}\vspace{-0.7\baselineskip}}]
        \item 	Dump truck location.
        \item 	Project location.
    \end{itemize}
 & 
    \begin{itemize}[nolistsep,leftmargin=*,before*={\mbox{}\vspace{-\baselineskip}},after*={\mbox{}\vspace{-0.7\baselineskip}}]
        \item 	Distance between truck and paver.
        \item 	Truck bed angle.
    \end{itemize} 
 & 
    \begin{itemize}[nolistsep,leftmargin=*,before*={\mbox{}\vspace{-\baselineskip}},after*={\mbox{}\vspace{-0.7\baselineskip}}]
        \item 	Screed speed.
        \item 	Mat thickness.
        \item 	Mat width.
        \item 	Mixture temperature.
    \end{itemize} 
 & 
    \begin{itemize}[nolistsep,leftmargin=*,before*={\mbox{}\vspace{-\baselineskip}},after*={\mbox{}\vspace{-0.7\baselineskip}}]
        \item 	Environmental condition.
        \item 	Mixture temperature.
        \item 	Mixture density.
    \end{itemize} \\

\textbf{Future} &
    \begin{itemize}[nolistsep,leftmargin=*,before*={\mbox{}\vspace{-\baselineskip}},after*={\mbox{}\vspace{-0.7\baselineskip}}]
        \item 	Time window for transportation.
    \end{itemize}
 & 
    \begin{itemize}[nolistsep,leftmargin=*,before*={\mbox{}\vspace{-\baselineskip}},after*={\mbox{}\vspace{-0.7\baselineskip}}]
        \item 	N/A.
    \end{itemize} 
 & 
    \begin{itemize}[nolistsep,leftmargin=*,before*={\mbox{}\vspace{-\baselineskip}},after*={\mbox{}\vspace{-0.7\baselineskip}}]
        \item 	Resurface thickness.
        \item 	Temperature window for paving.
    \end{itemize} 
 & 
    \begin{itemize}[nolistsep,leftmargin=*,before*={\mbox{}\vspace{-\baselineskip}},after*={\mbox{}\vspace{-0.7\baselineskip}}]
        \item 	Temperature window for compaction.
    \end{itemize} \\
\hline

\hline
\end{tabular}
\end{table}

Correspondingly, a major contribution from this research prospect will be the developed knowledge architecture for smart outdoor workplace. Unlike information architecture that aims at catering the existing and known information entities, knowledge architecture is supposed to deal with not only the existing but also the potential and future knowledge assets \cite{Gent_2008}. Thus, the developed knowledge architecture will naturally be extendable for generic smart workplace and other smart environments through broader research efforts and collaborations.

\subsection{Model-driven Simulation \& Optimization of Energy Consumption in Smart Outdoor Workplace}
Given the diverse workflows and the complex (possibly invisible) sensor network in outdoor workplace, it is impractical to monitor how individual sensors consume energy, not to mention their holistic energy consumption. Consequently, the modeling approach tends to be promising to relieve the challenges and complexity in this case, by abstracting real-world objects or processes that are difficult to observe or understand directly \cite{Referece_2018}. Furthermore, considering that a ``model is an abstraction of reality in the sense that it cannot represent all aspects of reality'' \cite{Rothenberg_1989}, multiple models will need to be weaved together to reflect the full scope of energy consumption in smart outdoor workplace, and each model addresses one of the aspects in the same context \cite{Mellor_Clark_2003}.

Based on the workplace breakdown analysis (e.g., Table \ref{table_breakdown}) together with the characteristics of the involved equipment, we can gradually develop energy consumption models for various outdoor workflow dynamics by including different energy concerns and factors, with regarding to different components, assumptions, application characteristics and environmental conditions. Take a known work on energy consumption of device-side data transmission  \cite{Segata_Bloessl_2014} as an example, if we are concerned with data size as the input factor only, a directly proportional relation between the energy consumption of a communication task and its data size can be assumed and modeled as:
\begin{equation}
\label{eqn:dataCommunication}
E_\mathit{device} = \lambda \cdot D
\end{equation}%
where $D$ is the data size, and $\lambda$ is a linear or quantile regression parameter that can be related to the employed access point technology.

When it comes to the holistic view, we suggest employing the agent-based modeling (ABM) technique \cite{Wilensky_Rand_2015} to model and integrate different workflow dynamics into a whole. As a powerful mechanism of explaining complex systems (e.g., social-ecological linkage and weather forecasting), ABM can encode the behavior of individual agents with simple rules to portray their interactions and the corresponding results. Note that the delivered agent-based model here will essentially be an abstract of the workflow-driven smart sensor network. By equipping individual agents with relevant energy consumption models and initializing suitable input parameters, it will be feasible to simulate and estimate the overall energy consumption within that predefined smart outdoor workplace. Furthermore, according to different simulation observations, it will also be possible to globally optimize energy consumption by adapting software concerns and/or adjusting workflow activities. ABM also provides the possibility of change different environmental  (weather, agent location, electronic interferences, devices failure)  and human behavior parameters (focus, skills), exploring the impact on the safety and energy consumption of the system.

From this research prospect, we expect a plug-and-play framework of modeling smart outdoor workplace with respect to the energy consumption.

\section{Conclusions and Future Work}
\label{sec:conclusion}

As a typical application scenario of outdoor workplace, road construction must take place in the natural and open areas.
Ideally, by making the communication between workers and devices/machines simpler and effective, and even by automating
some of the everyday tasks, a smart workplace implementation in this case will not only achieve the efficiency
and productivity of road construction, but also address its most crucial concern, namely safety \cite{SWA_2014}. However, we met various issues and difficulties  when we tried to equip the traditional transport ecosystem with the smart-workplace knowledge and technologies. 

To obtain more comprehensive lessons, we explored root causes of those issues and difficulties. Our efforts reveal that there are both theoretical confusions and technical challenges for implementing smart workplace, not to mention the unique characteristics of outdoor environments and activities. Therefore, driven by this special case of road construction and based on our exploratory  investigation, we conclude that it is worth paying dedicated attentions to the specific field ``smart outdoor workplace'' under the umbrella discipline of smart environment, as the ``outdoor workplace'' would cater more sophisticated workflows and more complex situations than the indoor environments (e.g., office, building, etc.).

Our future work will focus on the development of a small-scale knowledge architecture and a case study on the smart outdoor workplace for road construction. In the meantime, we will try to broadcast the dedicated idea of smart outdoor workplace in the whole community. There is no doubt that more efforts and collaborations are needed to solid this emerging topic.   

%
%
 \bibliographystyle{splncs03}
 \bibliography{smartcomRef}
\end{document}